%% file: main.tex
\documentclass[10pt,conference]{IEEEtran}

\usepackage{tabularx}
\usepackage[utf8]{inputenc}
\usepackage{multirow}
\usepackage{soul}
\usepackage{xcolor}
\usepackage{xspace}
\usepackage{url}
\usepackage{graphicx}
\usepackage{listings,multicol}
\usepackage[caption=false]{subfig}  
\usepackage[export]{adjustbox}
\usepackage{amsmath}
\usepackage{textcomp}
\usepackage{listings}
\usepackage{hyperref}
\usepackage{cleveref}
\usepackage{algorithm}
\usepackage{enumitem}
\usepackage{tcolorbox}
\usepackage{booktabs}
\usepackage{siunitx}
\usepackage{pifont}
\usepackage[noend]{algpseudocode}
\usepackage{balance}
\usepackage[noadjust]{cite}

\PassOptionsToPackage{svgnames}{xcolor}

\newcommand{\dataset}{\textsc{ConPlag}\xspace}
\newcommand{\stats}[1]{\textcolor{gray}{#1}}

\title{Towards a Dataset of Programming \\ Contest Plagiarism in Java}

\author{
 \IEEEauthorblockN{Evgeniy Slobodkin, Alexander Sadovnikov}
    \IEEEauthorblockA{\textit{Sirius.Courses, Moscow, Russia}}
    \IEEEauthorblockA{\{slobodkin.es, sadovnikov.av\}@talantiuspeh.ru}
}

\begin{document}

\maketitle

\begin{abstract}

In this paper, we describe and present the first dataset of source code plagiarism specifically aimed at contest plagiarism. The dataset contains 251 pairs of plagiarized solutions of competitive programming tasks in Java, as well as 660 non-plagiarized ones, however, the described approach can be used to extend the dataset in the future. Importantly, each pair comes in two versions: (a) ``raw'' and (b) with participants' repeated template code removed, allowing for evaluating tools in different settings. We used the collected dataset to compare the available source code plagiarism detection tools, including state-of-the-art ones, specifically in their ability to detect contest plagiarism. Our results indicate that the tools show significantly worse performance on the contest plagiarism because of the template code and the presence of other misleadingly similar code. Of the tested tools, token-based ones demonstrated the best performance in both variants of the dataset.

\end{abstract}

\input{sections/01-introduction}
\input{sections/02-background}
\input{sections/03-dataset}
\input{sections/04-analysis}
\input{sections/05-discussion}
\input{sections/06-threats}
\input{sections/07-conclusion}

\bibliographystyle{IEEEtran}
\balance
\bibliography{IEEEabrv,cite}

\end{document}

%% file: sections/01-introduction.tex
\section{Introduction}

In recent years, competitive programming contests have become increasingly popular~\cite{codeforces_report}. In such contests, participants solve different algorithmic problems, constrained by certain time and memory limits, and are judged by the speed of solving and the operation time of their solutions. 
Unfortunately, as the popularity of programming contests grows, so does the number of cases of \textit{plagiarism}~\cite{contest_warsaw2019}. 
To ensure the fairness of the contest and to enforce the rules of contest platforms, such cases need to be detected.

Unlike other \textit{semantic clones}~\cite{ain2019systematic}, plagiarized code often reuses particular popular transformations to hide the similarities~\cite{comparison2009}. Currently, there exist several tools based on different approaches to detect plagiarism in the source code~\cite{sherlock, sim1999, moss2003, dolos2022, jplag2003, bplag2021}, however, the majority of them are developed for detecting plagiarism in programming homeworks. 
At the same time, contest plagiarism has specific features that distinguish it from other kinds of plagiarism, the most important feature being the reuse of \textit{templates}. Template code in the context of competitive programming is the code that is written by the participant before the contest and copied between tasks in order to save time during the competition. 

Because of this, existing tools may perform worse on competitive plagiarism in comparison to other kinds. However, to the best of our knowledge, there is no single benchmark to compare the existing approaches in the setting of contest plagiarism. 
In this work, we strive to bridge the existing gaps in research, provide the community with the first version of curated dataset of specifically contest plagiarism, and compare available source-code plagiarism detection tools.

In the first part of our work, we curated the dataset. To do this, firstly, we selected 21 problems from the CodeForces platform~\cite{codeforces} and collected all 4,695 accepted Java submissions for them. We chose the Java language because it is the only programming language supported by the majority of plagiarism detection tools, thus allowing us to compare them in equal settings. For each problem, we constructed a set of all pairs from the submissions for this task. However, manually labelling a sample directly from all of the pairs will result in a lot of simple non-plagiarized examples, because the vast majority of submissions are not actually plagiarized. To overcome this problem, we decided to use the existing plagiarism detection tools based on different approaches~\cite{sherlock, sim1999, moss2003, dolos2022, jplag2003, bplag2021} to filter out \textit{trivially non-plagiarized} pairs.

The manual labeling demonstrated that five out of six tools sometimes give very low scores for actual plagiarism. Thus, we used the sixth tool, BPlag~\cite{bplag2021}, which demonstrated no plagiarized pairs below the similarity of 60\%. We filtered out all pairs below this threshold, leaving only ``interesting'' pairs, from which we selected a sample that was once again labeled manually. This resulted in a dataset called \dataset that contains 251 plagiarized pairs and 660 non-plagiarized ones.
Since an important feature of the contest submissions is template code, 
to have the opportunity to compare the tools in different settings, we manually removed templates from all the submissions, and share \dataset in two versions --- with and without them. \dataset is available online~\cite{dataset}.

In the second part of our work, we carry out the first pilot comparison of the existing approaches on both versions of \dataset. In this comparison, we used the same six main available tools as in filtering, since the final dataset only contained manually curated pairs: Sherlock~\cite{sherlock}, SIM~\cite{sim1999}, MOSS~\cite{moss2003}, Dolos~\cite{dolos2022}, JPlag~\cite{jplag2003}, and BPlag~\cite{bplag2021}. On the raw version of the dataset, the best results were obtained using token-based tools: JPlag, MOSS, and SIM. The analysis of the results demonstrates that some of their errors occurred due to the matching of the I/O code, similar in form between solutions. Also, unexpectedly low results were demonstrated by BPlag, the only graph-based algorithm. 
Finally, the approaches demonstrated better results when tested on the version of \dataset without templates, indicating that their removal is an integral part of developing further approaches.

We believe that \dataset and our pilot comparison is a valuable first step in solving the problem of plagiarism in programming contests that can be continued further. 

%% file: sections/02-background.tex
\section{Background}
\label{sec:background}

\subsubsection{Plagiarism in Programming Contests}

Source code plagiarism is a well-researched area of studies~\cite{marins2014survey}, however, the developed solutions are usually focused on finding plagiarism in homework assignments~\cite{novak2016academia, plugiarismreview2019},  
there are only a few works devoted to plagiarism in competitive programming~\cite{contest_bangladesh2021, contest_warsaw2019}. Even then, they are mainly focused on integrating a particular plagiarism detection tool into the online judging system, and not on comparing different existing tools. 

Unlike some other cases, source code in programming contests has its own specifics that directly relate to finding plagiarism in it. In addition to the usual plagiarism hiding techniques~\cite{comparison2009}, the solutions will have a lot of similar code that is natural for contests. Firstly, there is \textit{template} code, \textit{i.e.}, some common implementations of popular algorithms that the contestant copies into every solution. Secondly, there is code that will be almost the same in every solution to the given problem but not actually plagiarized, \textit{e.g.}, reading the input or printing the answer.
Currently, active research is underway on how to find similar code and reduce its influence on the similarity~\cite{common2016, common2020}.

The template code is particularly difficult to take into account, because it can be completely different for different contestants in both size and implementation. Also, functions from the template code can be actively used or ignored completely by the contestant, depending on a particular task, making it very difficult to automatically preprocess all submissions by simply removing all template code from them. It is clear that template code can easily become a weak spot for many algorithms, and this must be taken into account when using plagiarism detection tools on the contest code and when building a benchmark for their comparison.

\subsubsection{Source Code Plagiarism Detection Tools}

Many different tools were developed aimed at detecting source code plagiarism~\cite{plugiarismreview2019}. \textit{Text-based} algorithms treat a program as a simple text, without taking into account its programming language. The advantages of such approaches are language-independence and high performance~\cite{comparison2009}, however, this comes at the expense of lower accuracy. A popular text-based tool is Sherlock~\cite{sherlock}. This tool converts the file into a sequence of string tokens, hashes it, and extracts a subsample of hashes. To determine the similarity of two programs, Sherlock calculates the similarity of the sequences of hashes.

\textit{Token-based} algorithms run a specific lexer on the program and compare token streams. Such approaches are still fast but consider a deeper representation of the program. One of the earliest plagiarism detection tools, SIM~\cite{sim1999}, applies an algorithm for finding the maximum sequence alignment to the resulting token sequence of given programs. The similarity of two programs is then defined as their alignment score. Another popular token-based tool, JPlag~\cite{jplag2003}, defines the similarity as the percent of tokens from the first sequence that can be covered by tokens from the second one. A different token-based tool, MOSS~\cite{moss2003}, is based on comparing the fingerprints of programs. A fingerprint is constructed in three steps: (1) all the \textit{n}-grams for the token stream are built, (2) these \textit{n}-grams are hashed, and (3) to avoid comparing big sets of hashes, MOSS uses a \textit{winnowing} algorithm to select a certain subset of hashes for each program. The idea of winnowing has got popular, and several tools appeared that are based on it. One such tool is Dolos~\cite{dolos2022}. Unlike MOSS, Dolos is open-source, supports more programming languages, and provides powerful visualizations of the results.

Finally, \textit{graph-based} algorithms build a graph (usually, a program dependence graph) of the program, which shows the dependencies of the data within the program. To avoid the naive solving of an NP-hard problem, they use certain heuristics. For example, BPlag~\cite{bplag2021} uses the idea of a Greedy-String-Tiling algorithm to find similar parts in the graphs of two programs.

Overall, it can be seen that there exist a lot of approaches for finding plagiarism in the source code, however, given the specifics of programming contests, it is not clear how well they perform in such a setting. To evaluate the existing tools, to help researchers further improve them for competitive programming, as well as to provide a benchmark for future solutions, in this work, we aim to collect the first dedicated dataset of programming contest plagiarism.

\begin{figure*}[htbp]
\centering
    \includegraphics[width=\textwidth]{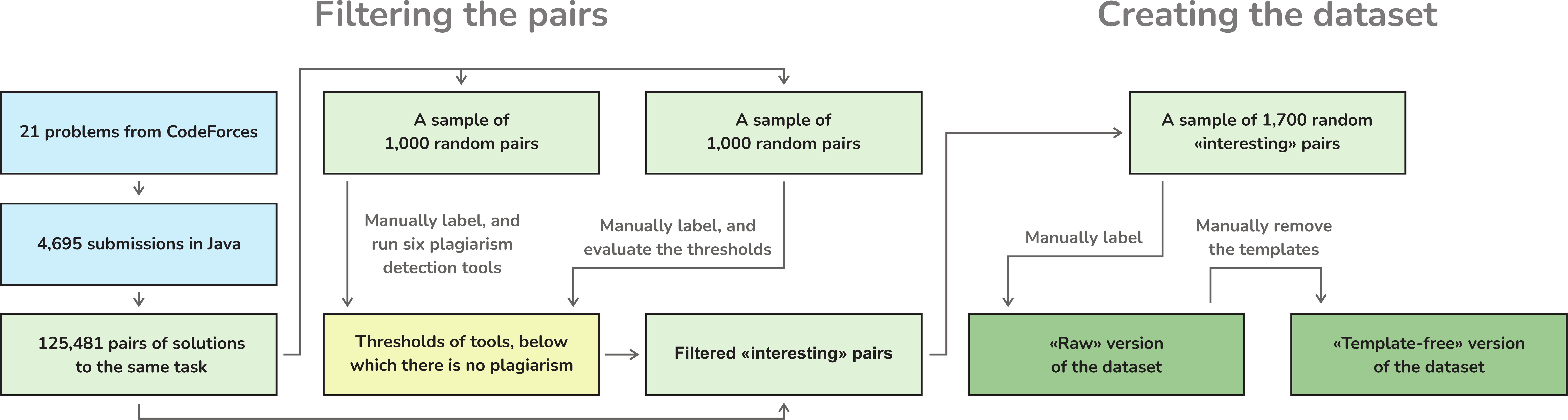}
    \centering
    \vspace{-0.5cm}
    \caption{The pipeline of the proposed approach for collecting the dataset.}
    \label{fig:pipeline}
    \vspace{-0.5cm}
\end{figure*}

%% file: sections/03-dataset.tex
\section{Dataset}
\label{sec:dataset}

When comparing source code plagiarism detection tools, researchers use (a) solutions to student assignments~\cite{jplag2003}, (b) synthetic augmentations of the programs~\cite{bplag2021}, or (c) manually changed code~\cite{sim1999}. 
To compare the tools in programming competitions, we decided to collect a dataset of pairs consisting exclusively of solutions to problems from real contests. 
The overall pipeline for collecting \dataset is presented in Figure~\ref{fig:pipeline}, let us now describe each step in greater detail.

\subsubsection{Gathering Data}

CodeForces~\cite{codeforces} is an online platform that hosts competitive programming contests.
The solution to the problem is a single-file program written in any popular language (C++, Python, Java, etc.). We chose CodeForces because it was used in previous research~\cite{majd2019code4bench, lobanov2023predicting}.

To create our dataset, we selected tasks on the platform based on the following two criteria: (a) the solutions must be large enough so that there are not many false positives when comparing them, and (b) there must be at least a hundred Java solutions for this task, otherwise the probability of plagiarism is very low. We collected a dataset of Java submissions because Java is the most popular language in plagiarism research and is the only language supported by all the tools available to us. 

Based on these criteria, we downloaded all successful Java submissions for 21 different problems and left only those that were successfully parsed by all the six studied tools~\cite{sherlock, sim1999, moss2003, dolos2022, jplag2003, bplag2021}, for a total of 4,695 submissions. Finally, for each problem, we built a set of all possible pairs of its solutions. This resulted in a total of 125,481 pairs.

\subsubsection{Excluding Trivial Non-plagiarized Code}

The constructed set of pairs is not feasible to manually label, and even more importantly, on the vast majority of these pairs, all plagiarism detection tools will return small values of similarity, which is true because the majority of contestants actually do not cheat. 
To overcome these difficulties, we decided to use the existing plagiarism detection tools, not to \textit{label} the data (since the idea is to obtain a manually-curated dataset) but to \textit{filter out} ``trivially'' non-plagiarized pairs. In particular, we used six plagiarism detection tools: Sherlock~\cite{sherlock}, SIM~\cite{sim1999}, MOSS~\cite{moss2003}, Dolos~\cite{dolos2022}, JPlag~\cite{jplag2003}, and BPlag~\cite{bplag2021}. The motivation behind using all the tools at once is to minimize the bias of the resulting dataset towards any one algorithm.

We took a sample of 1,000 random pairs and manually labeled them as either plagiarism or not. The labeling in this pilot work was carried out by the first author, who has 5 years of experience in competitive programming. In the future editions of the dataset, we plan to update it with more examples labeled by multiple experts. When the sample was labeled, we ran all the selected tools on these pairs. This way, for each tool, the \textit{minimum similarity} for the plagiarized pair was found. This threshold does not indicate that all the pairs above it are actually plagiarized, it merely indicates that \textit{all pairs below this threshold are not}. This resulted in 60\% for BPlag, 17\% for Dolos, and 0 for other four, meaning that there exists actual plagiarism, for which these tool show a score of 0. From these results, one can see that only BPlag can be used for effective filtering.

To ensure the reliability of the obtained threshold, we took another 1,000 random pairs, and the first author manually labeled them. Among the selected pairs, only 3 plagiarized pairs were detected, for which BPlag demonstrated less than 60\%, which indicates that this threshold allows us to successfully filter out simplistic cases.
Having obtained and evaluated the threshold, we ran BPlag on all the pairs and filtered out trivially non-plagiarized code. This removed approximately 57\% of the pairs, and left us only with ``interesting'' pairs.

\subsubsection{Building the Dataset of Pairs}

Next, we carried out the final manual labeling of the dataset. We took 1,700 pairs from the remaining set, and they were once again labeled by the first author. Overall, the labeling resulted in 251 cases of plagiarism and 660 cases of non-plagiarism. For the remaining 789 pairs, it was problematic to label them unambiguously, so we excluded such pairs.

\subsubsection{Dealing with Templates}

As discussed in Section~\ref{sec:background}, templates are an important and difficult feature of contest plagiarism. Therefore, for a more informative comparison, we decided to create a separate version of the dataset without the template code. Such a comparison would be further away from the real-world scenario, but would allow us to see which tools suffer the most from the templates.

After manually investigating many submissions, we found that all the template code can be generally classified into two groups: (a) fast input-output (I/O) and (b) the implementation of popular algorithms. We noticed that most of the algorithmic template code in the solutions is not used, so such code can be safely removed. However, the code for fast I/O is actually used by the contestant, so it can not be simply deleted, because some tools require the code to be compilable. In particular, this is important for graph-based tools like BPlag~\cite{bplag2021}, which requires the declaration of all the used functions. To deal with this, we generated a separate JAR-file with all such classes present, and in which BPlag could see the declarations of fast I/O methods. After this, we could safely remove the code responsible for fast I/O from each solution. 
The deletion of template code, both algorithmic and responsible for fast I/O, was done manually for each solution in each pair. 

\subsubsection{Dataset Characteristics}

Finally, we obtained a dataset of source code plagiarism in programming contests called \dataset that has a total of 911 pairs that were all manually labeled: 251 plagiarized pairs and 660 non-plagiarized pairs. Additionally, \dataset is presented in two different versions, meaning that there are two different versions of each solution: (a) \textbf{raw} --- each program in its original form, and (b) \textbf{template-free} --- with most template code removed.
\dataset is available online~\cite{dataset}, and comes with an easy-to-use CLI wrapping and detailed instructions. The evaluation of any tool can be done in two simple steps. Firstly, a special Python class should be created that will run the tool on each pair of the desired version of the dataset. Secondly, after the results are obtained, a separate script will compare them with the labeling and output the report, containing different metrics for the evaluated tools. 

%% file: sections/04-analysis.tex
\section{Comparing Different Techniques}
\label{sec:analysis}

The second part of our work consists in taking both versions of \dataset and comparing the existing tools on them. This can show us which tools work the best specifically in the setting of a programming contest, and also point the way to possible further improvements. 

\subsubsection{Metrics}

Firstly, we need to select the metrics for the comparison. As the basics, we can use \textit{precision} and \textit{recall}, two of the standard metrics used in software engineering research. At the same time, we need some other metric that would take into account the balance of precision and recall. One single main metric is necessary not only to make some conclusions about the performance of the tools, but also to be used for training the parameters of the tools.

To find the best trade-off between precision and recall, researchers typically use F1 score, \textit{i.e.}, the harmonic mean between their values. However, in our task, these two metrics are not actually completely equal. In real contests, declaring the solution as being plagiarized is a very serious decision, since this inevitably leads to severe consequences. Therefore, any cases of automatic plagiarism detection will be necessarily checked manually, and so finding all potential cases is more important. Luckily, F-score is actually a more general metric:

$$F_\beta = (1 + \beta^2) \cdot \frac{Precision \cdot Recall}{\beta^2 \cdot Precision + Recall}$$

\noindent where $\beta =$ 1 corresponds to the traditional F1 metric, $\beta <$ 1 favors precision, and $\beta >$ 1 favors recall. Since in our case, the recall is more important, we decided to use $\beta =$ 1.5. Further on, we will denote this metric as $F_{1.5}$.

\subsubsection{Approaches}

In our comparison, we used the same six main tools that were tested for the preliminary filtering of the dataset and that were described in Section~\ref{sec:background}. Since these tools are of different types, this comparison can show us which perform the best for programming contests and how much the template code affects them. 

Each tool under investigation has its own parameters that can be tuned. A lot of research in the adjacent area of clone detection indicates that the default parameters of clone detection tools are not actually the best ones~\cite{wang2013searching, golubev2021multi}. Therefore, to ensure the best performance for each tool, its different configurations were considered. The chosen configurations are similar to those used in the recent comparison~\cite{dolos2022}, there was a total of more than 400 configurations. You can find their full list, as well as the best ones, in the replication package~\cite{dataset}.

\subsubsection{Methodology}

In order to choose the best configuration for each tool, the dataset was divided into train (230 pairs) and test (681 pairs) sets. This division of the dataset is motivated by the fact that with the smaller size of the test dataset, it will be more difficult to find significant differences in the performance of the source code plagiarism detection tools.

Firstly, we ran all the tools on the train set. This means running each tool in each of the configurations with all possible similarity thresholds from the range \texttt{[0, 100]}. The best configuration of each tool was deemed the one that reached the maximum value of $F_{1.5}$. This search was carried out independently for two versions of the dataset, obtaining the best configurations separately for them. 

Then, we compared the best configurations of tools on the test set, also separately for the ``raw'' and the ``template-free'' versions of the dataset, thus allowing us to see the differences between them. We used bootstrapping~\cite{diciccio1996bootstrap} to build confidence intervals for the used metrics. Specifically, for each tool, we built 10,000 sub-samples with the same size as the test set, ran the tool on all of them, and then used the distribution of the metric to build a 95\% confidence interval. 

\subsubsection{Results}

Table~\ref{table:results} shows the results for both versions of \dataset. Firstly, we can see that the best results on both datasets are demonstrated by token-based tools. On the raw version of the dataset, JPlag demonstrated the best results in terms of the $F_{1.5}$ metric (0.77), however, MOSS and SIM are close to it (0.72 and 0.72, respectively). On the template-free dataset, JPlag also won (0.80), and all the four token-based tools demonstrate good results. 
As for the text-based tool Sherlock, in both versions of the dataset, it demonstrated good recall (0.76 and 0.81, respectively), but very low precision (0.34 and 0.39, respectively). While recall is in general more important to us than precision, such an extreme difference ceases to be useful. Interestingly, the graph-based tool BPlag also demonstrated results worse than token-based tools ($F_{1.5}$ of 0.55 on the raw dataset). Our investigation showed that it is too sensitive for programming contests --- a lot of pairs that have large similar chunks of code get counted as plagiarized.

\begin{table}[t]
\centering
\begin{tabular}{c  c  c  c } 
 \toprule
 \textbf{Tool}     & \textbf{Precision}          & \textbf{Recall}            & $\mathbf{F_{1.5}}$      \\ \midrule\midrule
 \multicolumn{4}{c}{\textbf{Raw dataset}} \\
 \midrule
 \textbf{Sherlock}  &  0.34 \stats{($\pm 0.05$)} & 0.76 \stats{($\pm 0.06$)} & 0.55 \stats{($\pm 0.05$)}  \\
 \textbf{SIM}       &  0.69 \stats{($\pm 0.06$)} & 0.74 \stats{($\pm 0.06$)} & 0.72 \stats{($\pm 0.05$)}  \\
 \textbf{MOSS}      &  \textbf{0.77} \stats{($\pm 0.06$)} & 0.71 \stats{($\pm 0.06$)} & 0.72 \stats{($\pm 0.05$)}  \\
 \textbf{Dolos}     &  0.68 \stats{($\pm 0.06$)} & 0.65 \stats{($\pm 0.07$)} & 0.66 \stats{($\pm 0.06$)}  \\
 \textbf{JPlag}     &  0.66 \stats{($\pm 0.06$)} & \textbf{0.83} \stats{($\pm 0.05$)} & \textbf{0.77} \stats{($\pm 0.05$)}  \\
 \textbf{BPlag}     &  0.45 \stats{($\pm 0.06$)} & 0.61 \stats{($\pm 0.07$)} & 0.55 \stats{($\pm 0.06$)}   \\ \midrule
 \multicolumn{4}{c}{\textbf{Template-free dataset}} \\
 \midrule
 \textbf{Sherlock} &   0.39 \stats{($\pm 0.05$)} &  0.81 \stats{($\pm 0.05$)} & 0.60 \stats{($\pm 0.05$)} \\ 
 \textbf{SIM}      &   0.73 \stats{($\pm 0.06$)} &  0.75 \stats{($\pm 0.06$)} & 0.74 \stats{($\pm 0.05$)} \\
 \textbf{MOSS}     &   0.66 \stats{($\pm 0.06$)} &  0.81 \stats{($\pm 0.06$)} & 0.75 \stats{($\pm 0.05$)} \\
 \textbf{Dolos}    &   0.72 \stats{($\pm 0.06$)} &  0.83 \stats{($\pm 0.05$)} & 0.79 \stats{($\pm 0.04$)} \\
 \textbf{JPlag}    &   \textbf{0.75} \stats{($\pm 0.06$)} &  0.83 \stats{($\pm 0.05$)} & \textbf{0.80} \stats{($\pm 0.04$)} \\
 \textbf{BPlag}    &   0.52 \stats{($\pm 0.06$)} &  \textbf{0.87} \stats{($\pm 0.05$)} & 0.72 \stats{($\pm 0.04$)} \\ \bottomrule
\end{tabular}
\vspace{0.2cm}
\caption{Results of comparing the best configurations of tools on the test set of \dataset, in two versions. Numbers in the brackets indicate 95\% confidence intervals.}
\vspace{-0.9cm}
\label{table:results}
\end{table}

Finally, the results show that the removal of the template code increases the performance of the studied tools. The difference is especially noticeable for Dolos (from $F_{1.5}$ of 0.66 to 0.79) and BPlag (from $F_{1.5}$ of 0.55 to 0.72). Without templates, both Dolos and BPlag greatly improve their recall, because they get less triggered by large similar pieces of code. These results indicate that the presence of the template code is indeed a serious challenge in correctly detecting plagiarism in programming contests.

%% file: sections/05-discussion.tex
\subsubsection{Discussion \& Implications}
\label{sec:discussion}

\textbf{Performance of tools.} The study of the existing tools on \dataset demonstrated that different tools show very different results. The most prominent result is that the graph-based tool BPlag turned out to be too sensitive for contest data, and demonstrated worse performance than simpler token-based tools. However, BPlag was able to detected certain non-trivial plagiarism cases that other tools missed. Future approaches might benefit from combining different kinds of code representation.

\textbf{Template code.} Another noticeable and expected result is that the performance of the tools improved on the version of the dataset without templates. For some tools, this change was very significant. This means that the automatic removal of templates can be considered as one of the most important tasks that can turn state-of-the-art plagiarism detection techniques into state-of-the-practice tools employed in real contests. 

\textbf{Repeating code.} Apart from the template code, solutions to the same task might have general repeated code, like reading or writing. A very important step in improving the quality of plagiarism detectors for contest code is being able to pay less attention to such code and more --- to the actual solution. A promising direction for further research would be automized ways of processing several solutions to the same task and marking the similar parts of the solution.

\textbf{Metric.} A traditional approach of locking a single threshold for the tool may not represent the real-life tool usage, where experts usually run the tool for each problem, get a list of pairs sorted by its similarity, and check pairs from the top of this list. This can be more representative when the tool has its own threshold for each problem. Ranking metrics from the information retrieval field might be a possible area of research. 

%% file: sections/06-threats.tex
\section{Limitations \& Future Work}
\label{sec:ttv}

\subsubsection{Language} Our dataset is based on the Java language, even though it is not the most prominent one in competitive programming. The choice of the language was dictated by the fact that it is the only one supported by all the prominent tools for detecting source code plagiarism. Future techniques, developed specifically for programming contests, may target more programming languages, such as C++. Our dataset cannot claim generalizability, however, the described approach can be used to collect similar datasets for other languages.

\subsubsection{Labeling} The labeling of the dataset was carried out by the first author of the paper and cannot be perfect. The author has a lot of experience with competitive programming, and we used several rounds of labeling to mitigate possible biases. At the same time, we view \dataset as the first step towards a fully fledged benchmark, and consider broadening it in the future as the main next step. For this reason, our major future plan is to conduct an extensive labeling with multiple experts.

%% file: sections/07-conclusion.tex
\section{Conclusion}
\label{sec:conclusion}

In this paper, we collected \dataset~\cite{dataset}, the first dataset of source code plagiarism consisting specifically of solutions to competitive programming contest tasks, and used this dataset to compare the existing plagiarism detection tools. The manually labeled dataset contains 251 pairs of plagiarized solutions and 660 pairs of non-plagiarized ones, and comes in two versions: ``raw'', as well as with removed templates.
The comparison of the existing tools for detecting software plagiarism showed that token-based tools demonstrate the best results, and that the presence of templates hinders the performance of all the tools.

%% file: main.bbl
\begin{thebibliography}{10}
\providecommand{\url}[1]{#1}
\csname url@samestyle\endcsname
\providecommand{\newblock}{\relax}
\providecommand{\bibinfo}[2]{#2}
\providecommand{\BIBentrySTDinterwordspacing}{\spaceskip=0pt\relax}
\providecommand{\BIBentryALTinterwordstretchfactor}{4}
\providecommand{\BIBentryALTinterwordspacing}{\spaceskip=\fontdimen2\font plus
\BIBentryALTinterwordstretchfactor\fontdimen3\font minus
  \fontdimen4\font\relax}
\providecommand{\BIBforeignlanguage}[2]{{%
\expandafter\ifx\csname l@#1\endcsname\relax
\typeout{** WARNING: IEEEtran.bst: No hyphenation pattern has been}%
\typeout{** loaded for the language `#1'. Using the pattern for}%
\typeout{** the default language instead.}%
\else
\language=\csname l@#1\endcsname
\fi
#2}}
\providecommand{\BIBdecl}{\relax}
\BIBdecl

\bibitem{codeforces_report}
CodeForces, ``{Results of 2020 (Annual Report)},''
  \url{https://codeforces.com/blog/entry/89502}, [Accessed 10-March-2023].

\bibitem{contest_warsaw2019}
Z.~Gniazdowski and M.~Boniecki, ``Detection of a source code plagiarism in a
  student programming competition,'' \emph{Zeszyty Naukowe Warszawskiej
  Wy{\.z}szej Szko{\l}y Informatyki}, vol.~13, no.~21, pp. 74--94, 2019.

\bibitem{ain2019systematic}
Q.~U. Ain, W.~H. Butt, M.~W. Anwar, F.~Azam, and B.~Maqbool, ``A systematic
  review on code clone detection,'' \emph{IEEE access}, vol.~7, pp.
  86\,121--86\,144, 2019.

\bibitem{comparison2009}
C.~Roy, J.~Cordy, and R.~Koschke, ``Comparison and evaluation of code clone
  detection techniques and tools: A qualitative approach,'' \emph{Science of
  Computer Programming}, vol.~74, pp. 470--495, 2009.

\bibitem{sherlock}
R.~Pike, ``{Sherlock Plagiarism Detector},''
  \url{https://web.archive.org/web/20180219024142/http://web.it.usyd.edu.au/~scilect/sherlock/},
  [Online. Accessed 10-March-2023].

\bibitem{sim1999}
D.~Gitchell and N.~Tran, ``Sim: a utility for detecting similarity in computer
  programs,'' \emph{ACM Sigcse Bulletin}, vol.~31, no.~1, pp. 266--270, 1999.

\bibitem{moss2003}
S.~Schleimer, D.~Wilkerson, and A.~Aiken, ``Winnowing: Local algorithms for
  document fingerprinting,'' \emph{Proceedings of the ACM SIGMOD ICMD},
  vol.~10, pp. 76--85, 2003.

\bibitem{dolos2022}
R.~Maertens, C.~Van~Petegem, N.~Strijbol, T.~Baeyens, A.~C. Jacobs, P.~Dawyndt,
  and B.~Mesuere, ``Dolos: Language-agnostic plagiarism detection in source
  code,'' \emph{Journal of Computer Assisted Learning}, vol.~38, no.~4, pp.
  1046--1061, 2022.

\bibitem{jplag2003}
L.~Prechelt, G.~Malpohl, M.~Philippsen \emph{et~al.}, ``Finding plagiarisms
  among a set of programs with {JPlag},'' \emph{J. Univers. Comput. Sci.},
  vol.~8, no.~11, p. 1016, 2002.

\bibitem{bplag2021}
H.~Cheers, Y.~Lin, and S.~P. Smith, ``Academic source code plagiarism detection
  by measuring program behavioral similarity,'' \emph{IEEE Access}, vol.~9, pp.
  50\,391--50\,412, 2021.

\bibitem{codeforces}
CodeForces, ``{Programming Contest Platform},''
  \url{https://codeforces.com/?locale=en}, [Accessed 10-March-2023].

\bibitem{dataset}
E.~Slobodkin and A.~Sadovnikov, ``{Dataset of Programming Contest Plagiarism in
  Java},'' \url{https://doi.org/10.5281/zenodo.7332490}, [Accessed
  10-March-2023].

\bibitem{marins2014survey}
V.~Martins, D.~Fonte, P.~Rangel~Henriques, and D.~da~Cruz, ``Plagiarism
  detection: A tool survey and comparison,'' \emph{OpenAccess Series in
  Informatics}, vol.~38, pp. 143--158, 2014.

\bibitem{novak2016academia}
M.~Novak, ``Review of source-code plagiarism detection in academia,'' in
  \emph{2016 39th International convention on information and communication
  technology, electronics and microelectronics (MIPRO)}.\hskip 1em plus 0.5em
  minus 0.4em\relax IEEE, 2016, pp. 796--801.

\bibitem{plugiarismreview2019}
O.~Karnalim, W.~Chivers \emph{et~al.}, ``Similarity detection techniques for
  academic source code plagiarism and collusion: a review,'' in \emph{2019 IEEE
  ICTALE (TALE)}, 2019, pp. 1--8.

\bibitem{contest_bangladesh2021}
F.~Iffath, A.~Kayes, M.~T. Rahman, J.~Ferdows, M.~S. Arefin, and M.~S. Hossain,
  ``Online judging platform utilizing dynamic plagiarism detection
  facilities,'' \emph{Computers}, vol.~10, no.~4, p.~47, 2021.

\bibitem{common2016}
C.~Domin, H.~Pohl, and M.~Krause, ``Improving plagiarism detection in coding
  assignments by dynamic removal of common ground,'' 2016, pp. 1173--1179.

\bibitem{common2020}
O.~Karnalim, J.~Sheard, I.~Dema, A.~Karkare, J.~Leinonen, M.~Liut, and
  R.~McCauley, ``Choosing code segments to exclude from code similarity
  detection,'' in \emph{Proceedings of the Working Group Reports on Innovation
  and Technology in Computer Science Education}, 2020, pp. 1--19.

\bibitem{majd2019code4bench}
A.~Majd, M.~Vahidi-Asl, A.~Khalilian, A.~Baraani-Dastjerdi, and B.~Zamani,
  ``Code4bench: A multidimensional benchmark of codeforces data for different
  program analysis techniques,'' \emph{Journal of Computer Languages}, vol.~53,
  pp. 38--52, 2019.

\bibitem{lobanov2023predicting}
A.~Lobanov, E.~Bogomolov, Y.~Golubev, M.~Mirzayanov, and T.~Bryksin,
  ``Predicting tags for programming tasks by combining textual and source code
  data,'' \emph{arXiv preprint arXiv:2301.04597}, 2023.

\bibitem{wang2013searching}
T.~Wang, M.~Harman, Y.~Jia, and J.~Krinke, ``Searching for better
  configurations: a rigorous approach to clone evaluation,'' in
  \emph{Proceedings of the 2013 9th Joint Meeting on Foundations of Software
  Engineering}, 2013, pp. 455--465.

\bibitem{golubev2021multi}
Y.~Golubev, V.~Poletansky, N.~Povarov, and T.~Bryksin, ``Multi-threshold
  token-based code clone detection,'' in \emph{2021 IEEE International
  Conference on Software Analysis, Evolution and Reengineering (SANER)}, 2021,
  pp. 496--500.

\bibitem{diciccio1996bootstrap}
T.~J. DiCiccio and B.~Efron, ``{Bootstrap confidence intervals},''
  \emph{Statistical Science}, vol.~11, no.~3, pp. 189 -- 228, 1996.

\end{thebibliography}
